\documentclass[twocolumn,prb,superscriptaddress,showpacs,citeautoscript,floatfix]{revtex4}
\usepackage{amsfonts}
\usepackage{amssymb}
\usepackage{amsmath}
\usepackage{graphicx}
\usepackage{dcolumn}   % column centering
\usepackage{bm}        % bold math
\usepackage{flafter}
\usepackage[dvips]{color}

\begin{document}

%\draft

\title{Probing Spin-Charge Relation by Magnetoconductance in One-Dimensional Polymer Nanofibers}

\date{\today}

\author{A. Choi}
\affiliation{Department of Physics and Astronomy, Seoul National
University, 599 Gwanak-ro, Gwanak-gu, Seoul 151-747, Korea}
\affiliation{WCU Flexible Nanosystems, Korea University, Seoul 136-713, Korea}
\author{K. H. Kim}
\affiliation{Department of Physics and Astronomy, Seoul National
University, 599 Gwanak-ro, Gwanak-gu, Seoul 151-747, Korea}
\author{S. J. Hong}
\affiliation{Department of Physics and Astronomy, Seoul National
University, 599 Gwanak-ro, Gwanak-gu, Seoul 151-747, Korea}
\author{M. Goh}
\affiliation{Department of Polymer chemistry, Kyoto University, Kyoto 615-8510, Japan}
\affiliation{Institute of Advanced Composite Materials, KIST, Jeollabuk-do 565-902, Korea}
\author{K. Akagi}
\affiliation{Department of Polymer chemistry, Kyoto University, Kyoto 615-8510, Japan}
\author{R. B. Kaner}
\affiliation{Department of Chemistry and Biochemistry, UCLA, Los Angeles, CA 90095, USA}
\author{N. N. Kirova}
\affiliation{CNRS-UMR 8502, Batiment 510, Universite Paris-sud, 91405 Orsay Cedex, France}
\author{S. A. Brazovskii}
\affiliation{LPTMS-CNRS, Batiment 100, Universite Paris-Sud, 91405 Orsay Cedex, France}
\author{A. T. Johnson}
\affiliation{Nano-Bio Interface Center, University of Pennsylvania, Philadelphia, PA 19104, USA}
\affiliation{Department of Physics and Astronomy, University of Pennsylvania, Philadelphia, PA 19104, USA}
\author{D. A. Bonnell}
\affiliation{Nano-Bio Interface Center, University of Pennsylvania, Philadelphia, PA 19104, USA}
\author{E. J. Mele}
\affiliation{Department of Physics and Astronomy, University of Pennsylvania, Philadelphia, PA 19104, USA}
\author{Y. W. Park}
\email{ywpark@phya.snu.ac.kr} \affiliation{Department of Physics and Astronomy, Seoul National
University, 599 Gwanak-ro, Gwanak-gu, Seoul 151-747, Korea}

\begin{abstract}
Polymer nanofibers are one-dimensional organic hydrocarbon systems containing conducting polymers where the non-linear local excitations such as solitons, polarons and bipolarons formed by the electron-phonon interaction were predicted. Magnetoconductance (MC) can simultaneously probe both the spin and charge of these mobile species and identify the effects of electron-electron interactions on these nonlinear excitations. Here we report our observations of a qualitatively different MC in polyacetylene (PA) and in polyaniline (PANI) and polythiophene (PT) nanofibers.  In PA the MC is essentially zero, but it is present in PANI and PT. The universal scaling behavior and the zero (finite)  MC in PA (PANI and PT) nanofibers provide evidence of Coulomb interactions between spinless charged solitons (interacting polarons which carry both spin and charge).
\end{abstract}

\smallskip


\maketitle

\section{Introduction}

Charge transport in conjugated polymers is due to the motion of charged species that are modified by an accompanying distortion of the lattice due to the electron-phonon interaction. These depend on the symmetry of the polymers \cite{1,2,3}, doping concentration \cite{4,5,6} and external force \cite{7,8}. Because each excitation has characteristic spin-charge relation, they can be probed by such as susceptibility measurements \cite{6,9} or optical spectroscopy measurements \cite{10,11,12}. Recently, the intersystem crossing of excitons in heterojunction systems was proved to be affected by a magnetic field \cite{13}. However, this does not directly give information about the spin-charge characteristics of the transport carriers. Films of conducting polymers have randomly entangled fibrillar morphology. In transport measurements, the orbital motion, i.e., the charge response to the Lorentz force in two- or three-dimensional (2D or 3D) polymer films, dominates the intrinsic spin response to the external magnetic field, $\mu$$_s$$\cdot$\textbf{H}$_{ext}$ \cite{14}. Thus, the intrinsic spin properties of the local excitations in conjugated polymer films have not been directly identified in charge transport experiments. Conversely, the polymer nanofibers have local crystalline bundles of one-dimensional (1D) chains along the fibers where the orbital motion is prohibited by their restricted dimensionality \cite{3,15}. Therefore, one can measure the intrinsic spin and charge response simultaneously by the magneto (probing spin) conductance (probing charge) measurements in 1-D polymer nanofibers. Furthermore, it is well known that the short-range electron-electron interaction becomes important in 1D systems, leading to a Luttinger-liquid (LL) state, and the effect of magnetic field to the spin-charge relation in LL have been studied recently \cite{16}. Assuming that the 1D transport is dominant in polymer nanofibers, i.e., assuming that the disorder due to the complex chain packing within the individual polymer nanofiber is perturbative, the study of magnetoconductance (MC) can thus estimate the effect of Coulomb interaction between the non-linear local excitations formed by the electron-phonon interaction.

In this paper, by modulating the external electric and magnetic fields, the intrinsic spin-charge characteristics of different local excitations for three different polymer nanofibers formed from polyacetylene (PA), polyaniline (PANI) and polythiophene (PT) are investigated. Detailed materials and methods are available as Supplemental Material \cite{17}.

\section{Experiment}
\subsection{Synthesis}
 A vertically aligned PA film was synthesized in a homeotropically aligned nematic liquid- crystal solvent under a vertical magnetic field \cite{18,19}. The aligned PA film has aligned fibril structure that is free from entanglement. The aligned PA is very easily dispersed, even without the surfactant. Due to an entanglement-free fibril morphology, a single PA nanofiber with $\sim$20-$\mu$m length is separated from the film well by ultrasonication for a very short time within 30 minutes in N,N-dimethylformamide (DMF) without a surfactant. The diameter of a single PA nanofiber ranges from 10 to 90 nm; the average diameter is 30 nm. In these newly designed PA single fibers, the long length and highly ordered morphology are of great advantages in studying the transport properties.

The PANI nanofibers used in this work were synthesized by a rapid mixing reaction between a solution of monomer with $p$-phenylenediamine as an initiator and oxidant solution to produce long and rigid nanofibers \cite{20}. Because an anisotropic reaction field was built with the aid of longer dimers than the aniline monomer, longer one-dimensional nanofibers ($\sim$2-30 $\mu$m) relative to the nanofibers synthesized using the method without the dimers ($\sim$2 $\mu$m) could be grown. The diameter of a single PANI nanofiber ranged from 30 to 90 nm; the average diameter was 50 nm.

The PT nanofibers were also synthesized by the rapid mixing reaction method \cite{21}. Similar to polyaniline nanofiber growth, long oligomers composed of six thiophene units helped to synthesize less entangled, long polythiophene nanofibers of several micron length. The diameter of a single PT nanofiber ranged from 30 to 50 nm; the average diameter is 35 nm.

The conducting polymers used in this work can be treated as 1-D systems because of their small diameters. By controlling the diameter of conducting quantum wires, a crossover from a 1-D to a 3-D with a number of weakly interacting channels has been observed in a molybdenum selenide nanowire \cite{22}. Additionally, polyacetylene is a conjugated polymer in which the weakly coupled linear chains of CH units form a quasi-one-dimensional lattice. Namely, local crystallinity exists. Compared to the diameter of 1-D inorganic nanowires (typically less than 20 nm), the diameters of the polymer nanofibers studied in this work are on the same order of magnitude (10-90 nm); some nanofibers even exhibit smaller diameters. Additionally, we could not discern a diameter dependence of the polymer nanofibers in this range. Because the distance between neighboring chains is significant (larger than 4 \AA\cite{23}), interchain interactions are very weak compared with intrachain interactions \cite{3}. Therefore, the boundary between a 1-D and 3-D is expected to lie far beyond the diameter of the conducting polymers we used.

\subsection{Sample preparation and doping}
A small droplet of the suspension of dispersed polymer nanofibers was deposited on top of 2-$\mu$m spaced electrodes pre-patterned on a Si wafer by photo lithography. An isolated single fiber was identified under an optical microscope. Pt or Au was used as the electrode materials to prevent a chemical reaction with dopant materials. For PA and PT, iodine doping was performed in the gas phase by exploiting the difference in pressure up to the saturation level and monitoring the sample current at a fixed bias voltage. The doping was performed right before the transport measurements to minimize the aging effect. Although the PANI had been doped during the synthesis, the sample was exposed to HNO$_3$ fume for a minute right before the transport measurement to dope the polyaniline nanofiber once again.

\subsection{ $I$-$V$ characteristics and MC}
The MC was measured using a 14-T superconducting magnet at the Nano Transport Laboratory in Seoul National University and an 18-T superconducting magnet or 35-T resistive magnet at the National High Magnetic Field Laboratory (NHMFL) in Tallahassee, Florida. The MC measurements were performed using a two-probe geometry and a Keithley 6517A electrometer. The temperature was kept stable by controlling the pressure of $^4$He during the sweeping of the magnetic field below $T$ = 4.2 K.

\section{Results and Discussion}
\subsection{Magnetic-field dependence}
\begin{figure}
 \includegraphics[bb=10 18 235 533, width=0.36\textwidth]{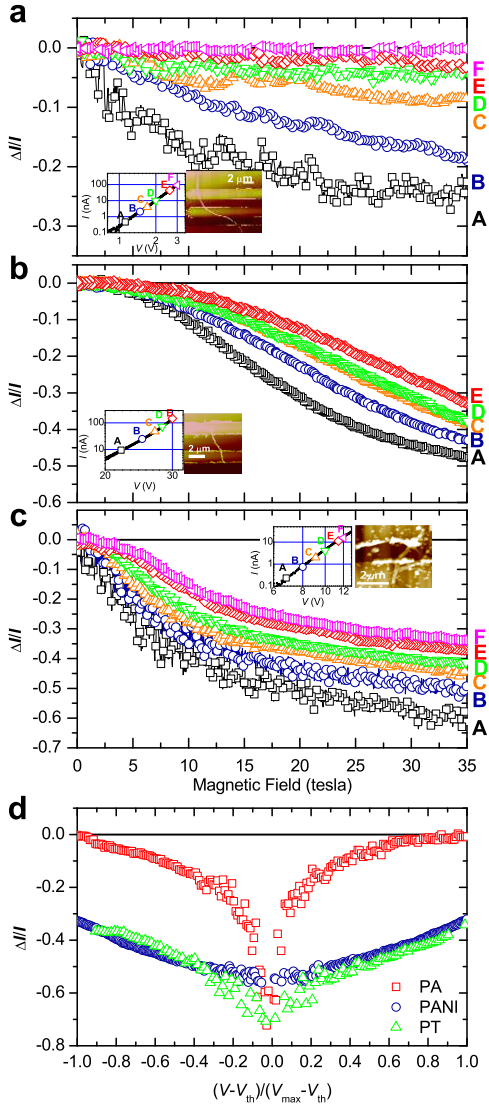}
 \caption{(Color) Electric field dependence of MC at 1.2 K. MC versus magnetic field $H$ for various fixed excitation voltages for (a) PA-1 nanofiber (A-F refer to each excitation voltage. A: $\sim$470 pA, 1.12 V; B: $\sim$2 nA, 1.48 V; C: $\sim$4 nA, 1.7 V; D: $\sim$10 nA, 2 V; E: $\sim$44 nA, 2.6 V; F: $\sim$90 nA, 3 V), (b) PANI-1 nanofiber (A: $\sim$9.5 nA, 22 V; B: $\sim$25 nA, 25 V; C: $\sim$48 nA, 27 V; D: $\sim$72 nA, 28.2 V; E: $\sim$147 nA, 30 V) and (c) PT-1 nanofiber (A: $\sim$230 pA, 6.8 V; B: $\sim$740 pA, 8 V; C: $\sim$2 nA, 9.1 V; D: $\sim$4 nA, 10 V; E: $\sim$11 nA, 11.4 V; F: $\sim$15 nA, 12 V). The $I$-$V$ curves in the inset of (a)-(c) are plotted on a log-log scale. (d) MC of each sample at 35 T versus the scaled excitation voltages, ($V_\mathrm{excitation}$-$V_\mathrm{threshold}$)/($V_\mathrm{applied~ maximum}$-$V_\mathrm{threshold}$). Here, $V_\mathrm{threshold}$ ($V_\mathrm{applied~maximum}$) = 0.86 V (3 V) for PA, 13.6 V (30 V) for PANI and 5.7 V (12 V) for PT, respectively.} \label{FIG.1}
\end{figure}

The MC, $\Delta$$I$/$I$$\equiv$[$I$($H$)-$I$(0)]/$I$(0), was obtained by measuring the current changes at fixed bias due to an applied magnetic field. Figure 1 shows the MC for (a) PA, (b) PANI and (c) PT nanofiber at various bias voltages as a function of magnetic field from 0 to 35 T at 1.2 K. For PA, the increasing magnitude of negative MC with magnetic field was observed at low bias voltage. However, the magnitude of MC decreases to zero with increasing bias voltage from a to f over the whole range of magnetic field. Once MC reaches zero at high voltage, a zero MC is maintained at the higher voltages. Furthermore, there is no orientation dependence in the MC of polymer nanofibers\cite{24}, the contribution of classical orbital motion of the charge carriers is ruled out. The anisotropic structure suppresses the orbital motion. Therefore, the magnetic field dependence of the conduction results from the spin state of charge carriers. This confirms that the spinless charged soliton de-confinement conduction in high electric field is responsible for the zero MC as proposed in our previous report \cite{24,25}. For PANI and PT nanofibers, on the other hand, the MC does not become zero in the highest applicable voltages, although it is also negative and its magnitudes tend to decrease with bias voltages. The continuous change of MC with respect to the scaled bias voltage at 35 T is shown in Fig. 1 (d) for the nanofibers. We will focus on the difference between the unique MC of PA nanofibers and the conventional MC of the PANI and PT nanofibers.

Figure 2 shows the color-map representation of the MC value at different $I$ and $V$ at $H$=14 T. The data guided by a solid line were measured for the same sample. The $I$-$V$ characteristics of PA, PANI and PT nanofibers at low temperature obey the power-law behavior $I$($V$)$\propto$V$^\mathrm{\beta}$. The MC of PANI and PT nanofibers never become zero in all cases; these are consistent with the results in Fig. 1. Furthermore, this confirms that the observed decrease in the magnitude of the MC of the PA nanofibers with increasing voltage and current is maintained over the entire $I$-$V$ range for all measured samples. In addition, we compared the PA, PANI and PT nanofibers of similar conductivity (current and bias levels, $\sim$230 pA and 6 V) in the Fig. S4 in Supplemental Material\cite{17}. Even though their conductivities are quantitatively similar, only the PA nanofiber shows magneto insensitive conductance. Therefore, distinct MC of PA does not come from the quantitative difference in conductivity or doping status, but from the qualitative difference in internal conduction mechanism.

\begin{figure}
\includegraphics[bb=13 14 235 336, width=0.36\textwidth]{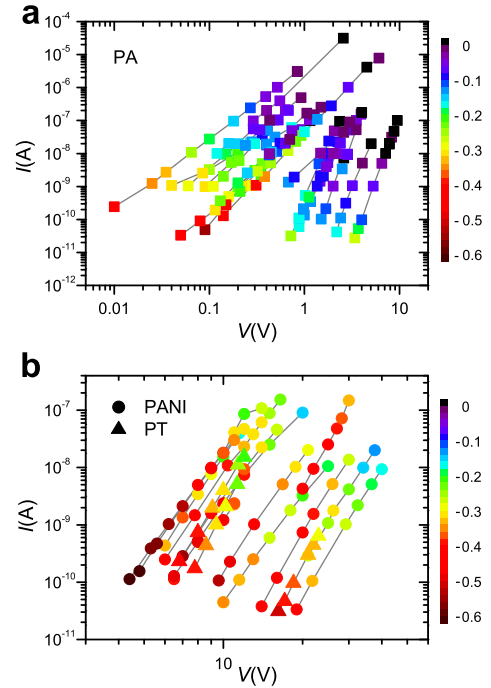}
\caption{(Color) Color map of MC of (a) the 22 PA nanofiber samples and (b) the 13 PANI nanofiber and 3 PT nanofiber samples in $I$-$V$ plot at magnetic field $H$=14 T and $T$=1.6 K. The color scales of (a) and (b) are identical.}\label{FIG.2}
\end{figure}

\subsection{Various Charge Carriers Conduction}
To explain the decreasing magnitude of MC with increasing bias voltages in PA nanofibers, we consider the two component currents, $I$=$I_\mathrm{e}$+$I_\mathrm{s}$, where $I_\mathrm{e}$ is the current from paramagnetic charge carriers that have spins and $I_\mathrm{s}$ is the current from spinless charged solitons. Magneto sensitive conduction is due to the motion of mobile charges of the paramagnetic species, which are polaron-defects along the conjugated chain \cite{26}. The negative MC from $I_\mathrm{e}$, appears at only at a low bias voltage when spinless solitons are bound in pairs. However, liberated solitons occur when the bias voltage is increased \cite{24}, so that the magneto insensitive $I_\mathrm{s}$ comprises the increasing proportion of the total current. Ultimately, the number of spinless solitons overwhelms that of the paramagnetic charge carriers at high electric field, eliminating the MC.

The negative MC of PA, PANI and PT in the low-electric-field regime can be explained with spin-dependent hopping model \cite{27}. The model predicts a negative sign for the MC and an $H^\mathrm{2}$ dependence for the magnetoresistance at low magnetic field. Although it is difficult to ascertain that the MC of PA possesses an $H^\mathrm{2}$ dependence, the MC of PANI and PT does indeed depend on $H^\mathrm{2}$ in low magnetic field.

In PA, the charged species are solitons which are domain walls separating two distinct although degenerate insulating states of the polymer. However, the interaction between neighboring chains in a fiber with local crystallinity can lift this degeneracy. This produces a confinement force that binds the soliton in pairs, estimated to be approximately 2 $\times$ 10$^\mathrm{5}$ eV/cm \cite{28, 29}. However, for the other known conjugated polymers, the degeneracy is lifted by the intrinsic structure of the chain because of the absence of a ground-state degeneracy on a single chain. The aromatic form is more stable by 0.351 eV/mol per ring than the quinoid form in PT \cite{30}. In PANI, the energy difference between the aromatic and quinoid form is roughly assumed to be the energy difference in polyparaphenylene, 0.439 eV/mol per ring \cite{30}. If we take 3.8 and 5.0 \AA as the repeating unit lengths in PT and PANI, respectively, we can estimate the corresponding confinement energy as 9.24 $\times$ 10$^\mathrm{6}$ eV/cm in PT and 8.78 $\times$ 10$^\mathrm{6}$ eV/cm in PANI. The confinement of spinless defects in PT and PANI is about two orders of magnitude stronger than the confinement of solitons in PA. Therefore, we could not observe the zero MC of spinless solitons in PANI and PT nanofibers. Also, the decrease of the magnitude of the MC in PANI and PT at a high electric field cannot possibly originate from the deconfinement of spinless defects even with the aid of bias voltage.

\subsection{Temperature Dependence of 1-D Conduction}

\begin{figure}
 \includegraphics[bb=22 28 235 594, width=0.36\textwidth]{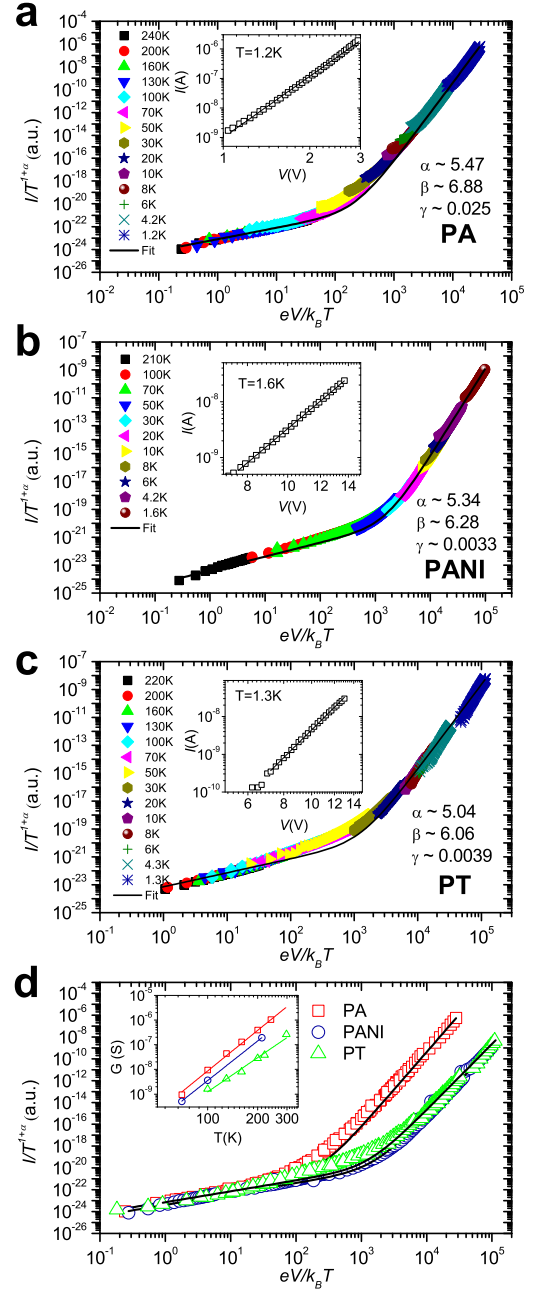}
 \caption{$I$/$T^{1+\alpha}$ versus eV/k$_B$T for (a) PA-2, (b) PANI-2, and (c) PT-1 nanofiber at different temperatures. The insets show low temperature $I$-$V$ characteristics which follow the power law $I$($V$)$\propto V^\beta$. (d) Collapsed $I$-$V$ curves for PA-2, PANI-2, and PT-1 nanofiber. The inset shows temperature-dependent conductance follows the power law $G$($T$)$\propto T^\alpha$.} \label{FIG.3}
\end{figure}

The fact that the zero MC in high electric field persists from $H$ = 0 to 35 T for PA nanofibers indicates that the intra-chain transport is maintained up to 35 T and higher magnetic field is necessary to deflect the charge carriers to the next chain forming closed-loop orbital motion. Therefore, we assume that the quasi-1D intra-chain charge transport is dominant up to 35 T in polymer nanofibers and adapt the LL model of the strictly 1D system to compare the conductivity results of polymer nanofibers, \cite{31,32,33} which treats the effects of electron-electron interactions in 1-D systems. Tunneling into a LL results in power-law relations of the form $G$($T$)$\propto T^\mathrm{\alpha}$ at low bias voltages (eV$\ll$ k$_B$T) and $I$($V$)$\propto$V$^\mathrm{\beta}$ at high bias voltages (eV $\gg$ k$_B$T), with $\mathrm{\alpha}=\mathrm{\beta}$+1 relation.

In addition, the $I$-$V$ curves in LL systems are predicted to be fitted by the general equation,
\begin{equation}
I = I_0 T^{1+\alpha}\cdot \sinh\left(\frac{\gamma eV} {k_B T}\right) \cdot \left| \Gamma \left( \frac{1+\beta} {2} + \gamma \frac{ieV} {k_B T} \right) \right| ^2 \label{equation1}
\end{equation}

where $\Gamma$ is the Gamma function, $I_0$ and $\gamma$ are constants, and $\alpha$ and $\beta$ correspond to the exponents derived from the slope of the fit line in the double-logarithm plot of $G$($T$) and $I$($V$). As shown in Fig. 3, all of the $I$-$V$ curves at different temperatures collapse well into a universal scaling curve for the polymer nanofibers. The $I$-$V$ curves at low temperatures with the threshold voltage, $V_\mathrm{t}$, are plotted above the region $V_\mathrm{t}$ (zero-current region was cut in the graph).

It is worth noting that the three different kinds of polymers synthesized by completely different methods show consistent behavior. In previous reports on the transport behavior of polymer nanostructures, $G$($T$) and $I$($V$) are described as a 3D variable-range hopping (VRH) model for PANI and polypyrrole tubes \cite{34,35} or a fluctuation-induced tunneling model for PA nanofibers \cite{36}. Additionally, the $G$($T$) of self-assembled PANI nanostructures are described using a 1D VRH model \cite{37}. Even the transport results for one kind of polymer nanostructure are not consistent with respect to the synthesis method, and there is no generally accepted model for the dominant transport mechanism. Although the diameters of polymer nanofibers and nanotubes are in the nanoscale range, exhibiting 1D transport behavior is another issue with respect to synthesis. Therefore, it is remarkable that similar universal scaling behavior was observed in three kinds of polymer nanofibers. The observed consistent universal scaling curves which are predicted in LL theory imply that the conduction in conducting polymer nanofibers is dominated by the Coulomb interactions between the charge carriers in 1D systems.

\subsection{Tunneling barriers}
The solid curves in Fig. 3 are the best fits to Eq. (1) with $\alpha$=5.47, $\beta$=6.88, and $\gamma$$\sim$0.025 for PA, $\alpha$=5.34, $\beta$=6.28, and $\gamma$$\sim$0.0033 for PANI, and $\alpha$=5.04, $\beta$=6.06, and $\gamma$$\sim$0.0039 for PT. The $\gamma$$^{-1}$ parameter has been interpreted as being related to the number of tunneling barriers that lie along the transport path in a LL \cite{31}. The corresponding $\gamma$$^{-1}$ values for PA, PANI and PT nanofibers are 40, 303 and 256, respectively. The $\gamma$$^{-1}$ values of the conducting polymer nanofibers in this work are 1-2 orders of magnitude larger than the values obtained in a clean LL system with a few tunnel junctions ($\sim$1-4) \cite{31,38,39}. However, the $\gamma$$^{-1}$ values of a quasi 1-D semiconducting polymer PBTTT film, which was treated using LL theory ($\sim$ a few hundred to a few thousand) \cite{40} were even larger than those of our polymer nanofibers. These unusually large values indicate the generality of the robust LL phenomenon in the presence of disorder which is present in our quasi 1-D system.

However, because of the relatively large values of the exponents and $\gamma$$^{-1}$, the clean LL theory may be modified. Aleshin $et~al$. \cite{33} suggested that the real transport mechanism in quasi 1-D polymer nanofibers obeys a single LL-like model valid for the different parts of a metallic polymer fiber separated by intermolecular junctions. The disorder such as the packing of finite length chains in the polymer nanofiber acts as the tunneling junction between the ends of two LL segments. This could result in larger exponents in $G$($T$) and $I$($V$) with respect to those in a clean LL. However, at low temperature, the measurement of $G$($T$) plot is not possible because of the blockade region below the $V_\mathrm{t}$; thus, other theories besides LL theory should be considered to understand charge transport in these materials. Aleshin $et~al$. \cite{41} explained the $I$-$V$ behavior of a PA nanofiber using Coulomb-blockade effects at low temperature with a crossover to LL-like behavior at high temperature. The structure consists of a 1-D array of small conducting regions separated by nanoscale barriers, resulting in a modified LL state at high temperature and a Coulomb-blockade state at low temperature.

The existence of the tunneling barriers between 1-D conducting regions does not mean that the tunneling is the main cause of the distinct behavior of MC. The conducting polymers and the electrodes in this work are not ferromagnetic materials. Therefore spin dependent tunneling might not affect to the magneto-transport of our systems. Also, tunneling through the barriers certainly exists in all the polymers. Therefore, the tunneling phenomena cannot explain the different behaviors of MC. If the number of barriers ($\gamma$$^{-1}$) is affecting the magneto transport, the MC can depend on the length or doping concentration. But we could not observe the non-zero MC at high electric field in long PA nanofibers or zero MC in short PANI nanofibers. And all of the PA samples with different conductivity (different doping level) show zero MC at high electric field (Fig. 2). Therefore we conclude that the tunneling phenomena and the quantitative difference of PA and other polymers are not the origin of the distinct MCs.

Through the magnetic field dependence of conducting polymer nanofibers, we investigated the distinct MC behavior involving various charge carriers with different spin-charge characteristics. Even with the tunneling barriers in the polymer nanofibers, the phenomenological robust 1-D nature of conjugated polymers and Coulomb interactions between charged quasi-particles has been observed in universal scaling behavior of conduction.

\section{Conclusions}
In conclusion, we have observed both of the spin dependent and spin independent conduction in conjugated polymer nanofibers by applying external electric and magnetic field. The various quasiparticles which have characteristic spin-charge relation play roles for conduction processes microscopically in different ways with respect to the kinds of polymers and applied excitation voltages considered. Liberated spinless charged solitons results zero MC in PA nanofibers at high electric field, while paramagnetic species appear itself by spin dependent transport at low electric field. On the other hand, spinless species in PANI and PT nanofibers cannot be liberated in the limit of maximally applicable electric field to these polymer nanofibers; effect of magnetic field on conductivity decreases but does not disappear in the measurement limit. Assuming that the effect of disorder due to complex chain packing in single polymer nanofiber is perturbative, so that the one dimensional charge transport is dominant, we analyze the $I$-$V$ characteristics of the three different polymer nanofibers with the modified LL. The observed distinct MCs and the universal scaling of PA (PANI and PT) nanofibers provide evidence of the Coulomb correlated spinless charged solitons (polarons which carry both spin and charge). The MC measurements for higher than 35 tesla would explore the effect of possible orbital deflection via interchain hopping transport at high enough magnetic field.

\section{Acknowledgments}

The high field MC was measured at the NHMFL in Tallahassee, Florida, USA. This work is supported by the Leading Foreign Research Institute Recruitment Program (0409-20100156) of NRF and the FPRD of BK21 through the MEST, Korea. Partial support for YWP was provided by the NBIC of U. Penn. AC also acknowledges support from World Class University program funded by the MEST through NRF of Korea
(R32-2008-000-10082-0).

\end{document}